\documentclass[11pt]{article}
\usepackage{geometry}                
\usepackage{graphicx}
\usepackage{amssymb}
\usepackage{amsmath}
\usepackage{cell}
\begin{document}

\title{\textbf{Protein structural variation in computational models and crystallographic data}}
\date{}

 \author{Dmitry A. Kondrashov$^1$$^*$, Adam W. Van Wynsberghe$^2$, Ryan M. Bannen$^1$, Qiang Cui$^3$,\\ and George N. Phillips, Jr.$^1$}

\maketitle
1: Department of Biochemistry; 2: Graduate Program in Biophysics; 3: Department of Chemistry and Theoretical Chemistry Institute, University of Wisconsin - Madison, Madison, WI 53706, USA. \\

*To whom correspondence should be addressed: dkon@biochem.wisc.edu. \\

Running title: Structural variation in computation and crystallography. \\


\section*{Summary} 
Normal mode analysis offers an efficient way of modeling the conformational flexibility of protein structures. Simple models defined by contact topology, known as elastic network models, have been used to model a variety of systems, but the validation is typically limited to individual modes for a single protein. We use anisotropic displacement parameters from crystallography to test the quality of prediction of both the magnitude and directionality of conformational variance. Normal modes from four simple elastic network model potentials and from the CHARMM forcefield are calculated for a data set of 83 diverse, ultrahigh resolution crystal structures. While all five potentials provide good predictions of the magnitude of flexibility, the methods that consider all atoms have a clear edge at prediction of directionality, and the CHARMM potential produces the best agreement.  The low-frequency modes from different potentials are similar, but those computed from the CHARMM potential show the greatest difference from the elastic network models. This was illustrated by computing the dynamic correlation matrices from different potentials for a PDZ domain structure. Comparison of normal mode results with anisotropic temperature factors opens the possibility of using ultrahigh resolution crystallographic data as a quantitative measure of molecular flexibility. The comprehensive evaluation demonstrates the costs and benefits of using normal mode potentials of varying complexity. Comparison of the dynamic correlation matrices suggests that a combination of topological and chemical potentials may help identify residues in which chemical forces make large contributions to intramolecular coupling. \\ \\
 
\small{Abbreviations: ENM, elastic network model; ANM, anisotropic network model; DNM, distance network model; BNM, block normal modes; ElN, ElNemo; HCA, harmonic C$\alpha$ potential; ADP, anisotropic displacement parameter}

\newpage
\section*{Introduction}
The native state of a protein is an ensemble of conformers, deviating to some extent from the average coordinates reported as the experimental structure. Knowledge of the static structure is not sufficient for understanding the functional mechanisms, which often depend on the flexibility of protein structures.  Experimental observation of conformational motion of biomolecules is becoming possible, thanks to experimental innovation, but remains a formidable challenge.  Crystals can be subjected to time-resolved experiments \cite{moffat2001-1569}, but the range of applications is limited to reactions that can be triggered by light or trapped by clever manipulations.  NMR spectroscopy can be used to determine both the structure and the dynamics of proteins \cite{lindorff2005-128}, but it is limited both by the maximum size of protein structures and by the difficulty of discrimination of slowly or quickly exchanging dynamics \cite{palmer2001-204}. Mass spectrometry coupled with hydrogen/deuterium exchange and proteolysis has been used to determine changes in the relative solvent accessibility of amide hydrogens \cite{lanman2004-181}, and single-molecule experiments using optical trapping have resulted in spectacular observations of the motion of motor proteins \cite{abbondanzieri2005-460}.  In general, direct measurement of molecular motion remains laborious and limited.

Computer simulations of biological macromolecules enable detailed explorations of the conformational ensemble near the native state \cite{karplus2005-6679}.  However, the computational cost of molecular dynamics with all-atom forcefields limits the accessible timescale of simulations, particularly of large molecular assemblies.  Thus, approximate methods, such as normal mode analysis (NMA), are often used to efficiently describe the allowed conformational ensemble of protein structures \cite{go1983-3696,brooks1983-6571,levitt1985-423}.  The decomposition into modes with different frequencies reduces the dimensionality of the problem, since a few lowest-frequency modes describe the most dominant directions of motion \cite{teodoro2003-617}. These global modes have been used to predict protein flexibility \cite{cui2004-345} and to study the mechanism of conformational transitions necessary for protein function \cite{ma1997-114}.  Simple coarse-grained potentials, such as Elastic Network Models, provide an  efficient description of a protein structure by connecting atoms or residues within a certain distance with identical harmonic potentials \cite{tirion1996-1905}.  Despite the extreme simplification, these models capture the basic topology of a structure and generate  predictions on the flexibility and preferred modes of motion of proteins that are in general agreement with experimental data \cite{bahar2005-586}.  

The study of protein conformational dynamics requires an interplay between experiment and computation. A readily available measure of conformational mobility is the Debye-Waller temperature factor, or B-factor, which models the variance in atomic position from the scattering data. It has been used as a source of information on protein flexibility for decades \cite{frauenfelder1979-558}, and as computational methodologies have matured, studies over large numbers of crystal structures have shown good agreement with computations,  specifically with ENM results \cite{kundu2002-723}.  While the classic B-factor has long been a routine parameter in protein structure refinement, until recently few crystal data sets contained sufficiently many observations (unique reflections) to allow determination of anisotropic displacement parameters (ADPs).  These parameters model the probability distribution of atomic positions as a Gaussian function with ellipsoidal contours, and have been shown to significantly improve the refinement statistics for crystal structures of biological macromolecules \cite{longhi1997-779,dauter1997-16065,esposito2000-713} at resolution better than 1.2 \AA. ADPs have been used in a few studies as a qualitative indicator of the directionality of prevalent motion in a protein structure \cite{wilson2000-1237}, but this source of experimental information has not been systematically exploited. 

The proliferation of various simple ENM-like models for macromolecular fluctuation begs the question of their relative fidelity and reliability, but no systematic comparison of the methods has been undertaken, to the best of our knowledge.  In a recent study, 98 highest resolution crystal structures in the PDB were used for systematic evaluation of prediction of the magnitude of motion in protein structures using an isotropic ENM model \cite{kondrashov2006-2760}.  In the present work, we compare the quality of prediction of the magnitude and direction of structural variance for the most commonly used anisotropic ENM potentials, and introduce a new one to better model different chemical interactions. Comparison between ADPs and computational variance matrices allows a quantitative evaluation of  the merits and drawbacks of different potentials. We also investigate the effect of the choice of potential on the global dynamic properties, such as the correlation matrix.

\section*{Results}
\subsection*{Analysis of crystallographic data} 
\begin{figure}[h]
\begin{center}
\vspace{.2in}
\centerline {
\includegraphics[width=5in]{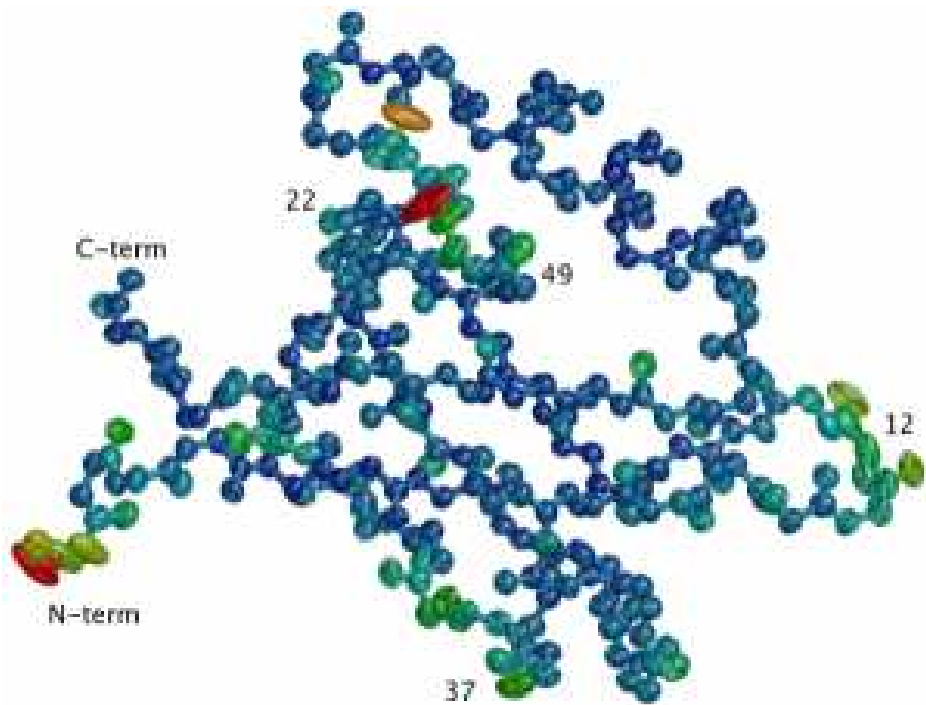}
}
\vspace{.2in}
\end{center}
\caption{Example of a high-resolution protein structure (PDB ID 1R6J) showing anisotropic temperature factors for backbone atoms. The ellipsoids represent 90\% probability volume of atomic position, with color varying from immobile (blue) to more mobile (red). Most of the backbone atoms are not mobile and isotropic, with a few loop residues (residue numbers labeled) showing clear directional preference in positional distribution.}
\label{fig:pdz_anis}
\end{figure}

\begin{figure}[h]
\begin{center}
\vspace{.2in}
\centerline {
\includegraphics[width=5in]{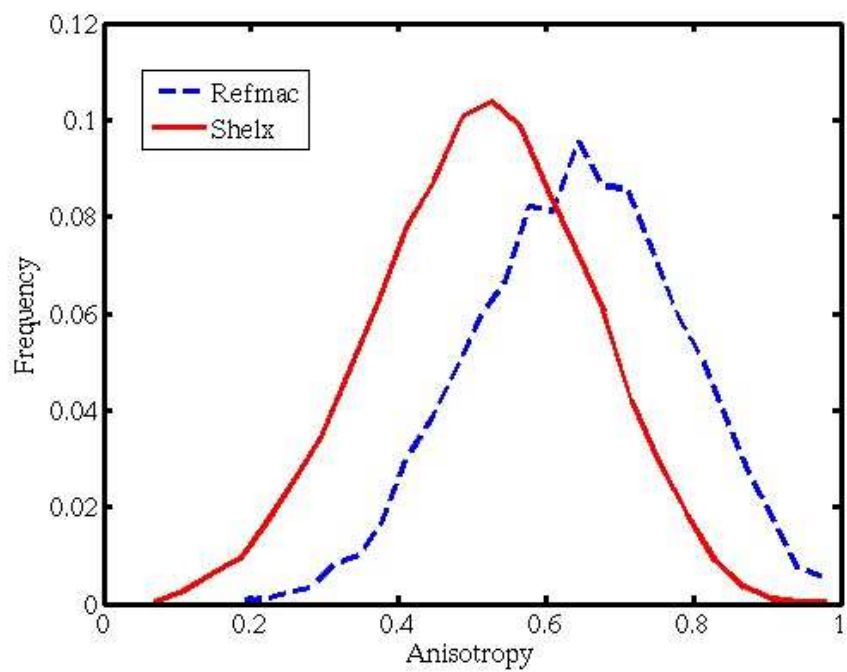}
}
\vspace{.2in}
\end{center}
\caption{Distributions of anisotropy parameters in structures refined with SHELX and Refmac software. The large difference is likely due to different default restraints on the anisotropic parameters in the two. }
\label{fig:anis_hist}
\end{figure}

The present study evaluates predictions of five coarse-grained normal mode potentials using a set of anisotropic displacement parameters from ultra-high resolution crystal structures. The Protein Data Bank \cite{berman2000-235} was searched for all X-ray crystal structures of proteins with chain length of at least 50 residues, with resolution at or beyond 1 \AA, with the restriction that the structures have less than 50\% sequence identity. 83 such structures were deposited with anisotropic displacement paremetrs, containing a total of 17763 protein residues.  Excluding those with disordered C$\alpha$ atoms or those involved in intermolecular crystal contacts, both of which have an effect on the ADP, left 12348 residues with usable ADPs.  The anisotropic displacement parameters are commonly represented as ellipsoids in crystal structures, as shown in Figure \ref{fig:pdz_anis}, and contain information about both the magnitude and the preferred direction of atomic variation in the crystal. The anisotropy of the ellipsoid, defined as the ratio of the smallest to the largest eigenvalue of ellipsoid matrix \cite{trueblood1996-770}, is a measure of deviation from spherical shape. We separated the structures by refinement software used, and found different distributions of anisotropy for the C$\alpha$ ADPs. 68 structures were refined using SHELX \cite{sheldrick1997-319}, and the remaining were determined 15 using Refmac from the CCP4 suite \cite{ccp41994-760}.  The C$\alpha$ ADPs in the Refmac set had a mean anisotropy of 0.64, compared with 0.51 for the SHELX set (Figure \ref{fig:anis_hist}), suggesting that the crystallographic restraints used in the two programs have significant effects on resulting ADPs.  Since sphere-like ellipsoids contain little directional information, a subset of ADPs with anisotropy of less than 0.5 was chosen, leaving 4642 ADPs to compare with the computational predictions of directionality of variance. 

\subsection*{Normal mode potentials} Elastic Network Models are dependent on the choice of cutoff distance \cite{tirion1996-1905}, which separates atom pairs deemed in contact from those which are not interacting. We introduce a new ENM method, called distance-based network model (DNM), an elastic network model with multiple force constants for atomic contacts, as described in Methods. It is clear that atom pairs closer than 2.3 \AA are covalently bound and thus have stronger interactions than those which are 5 \AA apart. To mimic the chemistry, several discrete distance ranges were defined and the force constants for each category are set to be the reciprocal of the total number of contacts in this range. Since the number of atomic contacts grows with distance, this ensures that interactions between atoms farther away are represented by weaker force constants than those in close proximity. We have shown recently that a similar strengthening of force constants between covalently bound residues resulted in greatly improved variance prediction quality for an isotropic ENM \cite{kondrashov2006-2760}, compared with the single force constant GNM \cite{bahar1997-173}. The atomic interactions are added up with the appropriate force constants for each residue, producing a residue-level model based on atomic interactions, with no additional free parameters, since the force constants are defined based on the contact matrices. The only parameter not defined from the structure is the maximum cutoff distance considered, and we optimized it by comparing prediction quality in calculations with a range of cutoffs from 5 \AA\ to 11 \AA. The analysis for agreement with magnitudes and directions of ADP ellipsoids with the model predictions is shown in Table 1 of SI. With the exception of the 5 \AA\ cutoff, the results were very similar, and 9 \AA\  was selected as the optimal cutoff distance.  

We also tested four existing normal mode potentials, three of the ENM variety, and one based on the CHARMM forcefield. The ElNemo method \cite{suhre2004-W610} depends on the cutoff distance between atoms, and the atomic contact matrices are combined into rigid-motion blocks on a residue level. We varied the cutoff distance from 5 to 11 \AA\ (Table 2 of SI), and found that the best results at 5 \AA, compared with the default value of 8 \AA.  The anisotropic network model (ANM) \cite{atilgan2001-505} depends on a cutoff distance between C$\alpha$ atoms, and we evaluated the results for a range from 10 to 16 \AA\ (see Table 3 of SI). The variation is also relatively small, but there is an opposite trend between quality of prediction of direction and magnitude of motion. The best cutoff for directionality prediction was at 10 \AA, while the best agreement in magnitude was with 16 \AA\ cutoff, in contrast to the previously used value of 13 \AA\ \cite{atilgan2001-505}. Normal modes were computed using an atomistic forcefield (CHARMM) with rigid-body blocks for residues, referred to as Block Normal Modes (BNM) \cite{li2002-2457}.  The inclusion of non-protein ligands and cofactors in the CHARMM force-field resulted in significant increases in quality of prediction, and thus all the non-protein residues for which CHARMM libraries could be found were added to the models. The last method used is the Harmonic C$\alpha$ potential (HCA) with distance-dependent force constant \cite{hinsen2000-25}, as implemented in Molecular Modeling Toolkit (MMTK) \cite{hinsen2000-79}. 

\subsection*{Comparison of crystallographic and computational variance}  Anisotropic covariance tensors were computed from 90 lowest frequency normal modes from 83 structures, as described in Methods, and fidelity of both magnitude and direction prediction was assessed. Magnitude prediction quality was measured by linear correlation between isotropic ADPs (B-factors) and the predicted isotropic variances over each structure. Two different measures were used for directional agreement, the absolute value of the dot product between the largest axes of the anisotropic ADPs (ellipsoids), and the real space correlation coefficient, as defined in Methods. These two measures were employed to compare pairs of corresponding residues, and the reported numbers are the statistics over all sufficiently anisotropic ellipsoids from all 83 structures. Table \ref{tab:results_summary} shows that prediction quality was markedly different for the magnitude and direction of motion. All the models had average isotropic correlations of 0.66-0.68, with the exception of 0.61 for HCA. On the other hand, there was considerable variation in the directional agreement of ADP ellipsoids. The two measures of directional agreement, the dot product and the real space correlation, largely showed the same trend, with HCA and ANM displaying relatively weak agreement, while ElN, DNM and BNM, show considerably higher prediction quality, with CHARMM-based BNM having an edge over the ENM methods. The mean absolute value of the dot product is easy to interpret as a measure of the angle between the preferred direction and in the experimental and computed ellipsoids. The average value of 0.65 for BNM corresponds to an angle of 48$^o$, while the average of 0.56 for ANM corresponds to an angle of 56$^o$, but this does not tell the whole story because it only compares one principal axis of the ellipsoids. The real space correlation is the volume fraction shared by two ellipsoids of unit volume, and this quantity varies appreciably from 0.52 for HCA to 0.61 for BNM. We tested the hypothesis that predictions agree no better than expected from a random uniform distribution of ellipsoid direction, for which the mean dot product is 0.5, and the mean real space correlation is 0.3 (when anisotropy is fixed at 0.5). Almost all of the structures with a reasonable sample of usable ADPs (with anisotropy $< 0.5$) showed better than random agreement  in real space correlation ($P<0.01$, see SI). For the worst-performing method, HCA, 11 structures did not meet this criterion, and only four had more than 10 sufficiently anisotropic ADPs, with the highest at 24. The results for the best-performing BNM method had only four structures where the null hypothesis could not be rejected, all of which had only 5 or fewer usable ADPs.

\subsection*{Effect of potential on global dynamic ensembles} 
\begin{figure}[h]
\begin{center}
\vspace{.2in}
\centerline {
\includegraphics[width=5in]{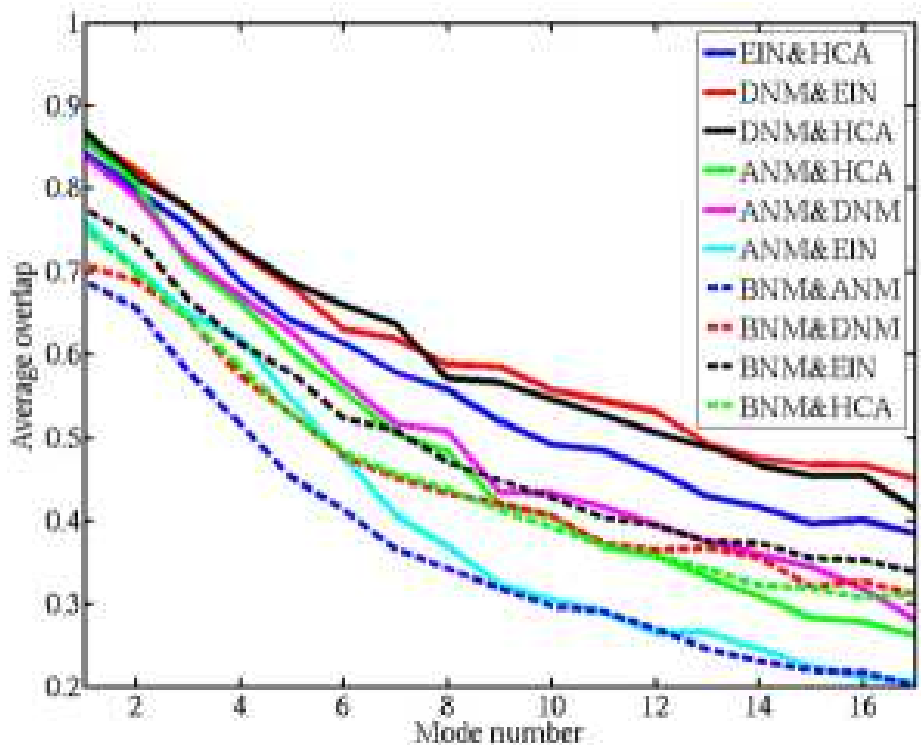}
}
\vspace{.2in}
\end{center}
\caption{Overlap scores for normal modes from different potentials averaged over all 83 structures. Each curve is a comparison between a pair of potentials over the 17 lowest frequency modes.  The solid curves compare different ENM-like potentials, while the dotted curves compare CHARMM-based BNM results with those from ENM potentials.}
\label{fig:mode_comp}
\end{figure}

Comparison of the normal modes produced by different methods revealed a clear distinction between the harmonic ENM models and the CHARMM forcefield BNM.  We used the modes computed from all 83 structures to investigate how the dynamic ensemble predictions depend on the use of the potential. The overlap measure described in Methods was computed for the 17 lowest-frequency modes, averaged over the entire data set, and plotted for all 10 pairs of methods in Figure \ref{fig:mode_comp}. The highest agreement was observed for the lowest-frequency modes, but the overlap measure dropped below 0.5, depending on the pair of methods, at some point in the first 15 modes. This demonstrated that the details of potential play a secondary role at lowest-frequency modes, which are dominated by the contact topology and shape of the molecular structure.  A second observation is the distinctiveness of modes derived from CHARMM-based BNM, which showed much lower overlap with ENM-based methods (dotted lines) than overlap among modes from ENM-type potentials (solid lines), with the single exception of the overlap between ANM and ElN. Since the latter is an all-atom potential, it is reasonable that it should be closer to chemistry-based BNM than to C$\alpha$-based ANM.  We tested the possibility that minimization of structures prior to BNM is responsible for the difference in BNM modes, by calculating DNM modes from the minimized structures. The resulting average overlap with BNM was 0.76 as opposed to 0.75 for BNM with DNM from unminimized structures, still much lower than DNM agreement with other methods. This suggests that the chemical information present in the all-atom CHARMM potential plays a role in determining the lowest-frequency modes, in addition to the topology of the structure.

\begin{figure}[!h]
\begin{center}
\vspace{.2in}
\centerline {
\includegraphics[width=6in]{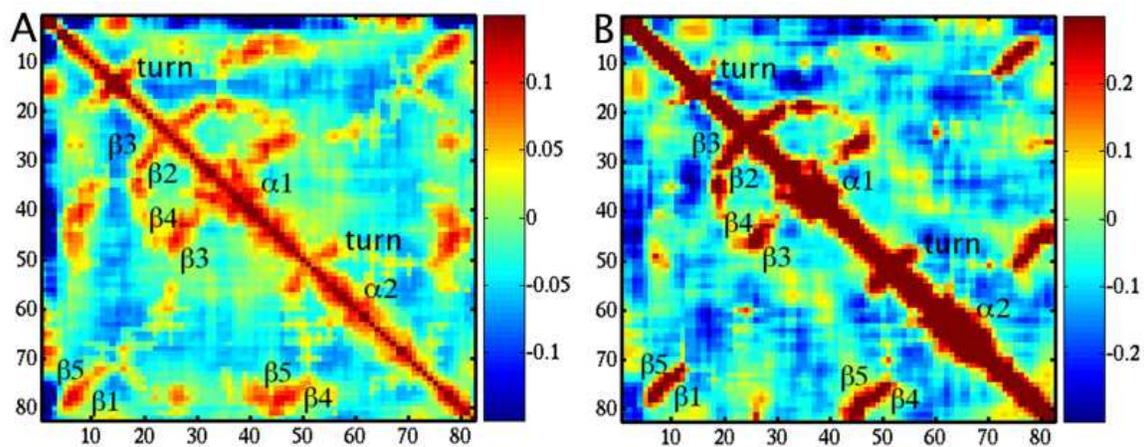}
}
\vspace{.2in}
\end{center}
\caption{Correlation matrices generated from normal mode analyses of PDZ domain (PDB ID 1R6J).  The plots show correlation between residues with indices shown on x and y axes, blue color indicating negative correlation and red signifying positive, with the range shown in the colorbars. Secondary structure elements are labeled in sequence order. A) Correlation from Anisotropic Network Model; B) from CHARMM-based Block Normal Modes. Note that the range in plot A is half that of B. }
\label{fig:corr_mat}
\end{figure}

To illustrate the differences between the chemical forcefield and ENM, we picked a small, well-studied structure from the data set, a PDZ2 domain from syntenin (PDB ID 1R6J) and computed the correlation matrices (see Methods) from the 90 low-frequency modes of ANM and BNM.  Figure \ref{fig:corr_mat} shows correlation matrices computed from ANM and BNM modes. In general, they look quite similar, with major features determined by the secondary structure elements: anti-parallel beta sheets appear as positive bands perpendicular to the diagonal, and the two helices result in a thickening of the diagonal band. While the pattern of secondary structures is clear in both potentials, there are evident differences. First, the magnitude of correlation is at least two times weaker in ANM (see the colorbar), and the secondary structure features are not as clear, due to the inclusion of residues as far as 16 \AA\ away.  Second, due to identical force constants for distant and proximal interactions, the diagonal band is considerably weaker in the ANM plot than in BNM, which has a more realistic representation of covalent bonds and other main-chain interactions. Both potentials capture the effect of gross topology, but the effects of specific chemistry are hidden in the fine details of the BNM correlation matrix.

\section*{Discussion}
We analyzed five different coarse-grained potentials used to model the conformational flexibility of protein structures. These were evaluated both by validation against experimental data and by comparison among the different potentials.  To our knowledge this is the first systematic attempt to use anisotropic displacement parameters to validate computational predictions, and it behooves us to note the challenges arising from using this data source. The reliability of ADPs has been tested before \cite{merritt1999-1997}, with good agreement in ellipsoid shape observed between independently determined structures of the same protein; we found the same to be true for structures of myoglobin in four different crystal forms (Kondrashov, et al, unpublished). This shows that ADPs are robust experimental parameters, and to minimize the noise contributions we used the highest resolution crystal structures available. However, quantitative comparison between ADPs and computational predictions are not straightforward, due to contributions of experimental noise, model error \cite{kuriyan1986-227}, rigid-body motion of the entire molecules \cite{kuriyan1991-2773}, and specifics of crystal environment, such as crystal contacts between copies of the protein packed in the lattice \cite{phillips1990-381} and collective lattice modes \cite{clarage1992-145}.  Further, the ADP represents the best fit of a Gaussian distribution to the electron density of an atom, but anharmonic and multimodal positional distributions are expected for protein atoms, especially in mobile regions, such as the surface. Only atoms with pronounced anisotropy are used for directional comparison, which tend to lie in mobile regions with poorer electron density (see Figure \ref{fig:pdz_anis}), and which are not adequately modeled with a single conformer \cite{depristo2004-831}. Thus it is likely that many atoms in our directional dataset are not adequately modeled by the Gaussian ADP model. Despite these caveats, our results show good agreement between the predicted and computed ADPs: for virtually all structures, the real space correlation between BNM predictions and ADPs is significantly better than the expectation from a uniform random variable. This suggests that the influence of the factors listed above is not sufficient to overwhelm the important contribution of intramolecular conformational flexibility. This is consistent with a recent comparison of MD simulations with crystallographic B-factors which estimated that rigid-body motions contribute only 20-30\% of total positional variance in B-factors \cite{meinhold2005-2554}. The agreement between computation and experiment serves to validate both the interpretation of the experimental data and the reliability of computational predictions.

In our analysis we combined multiple low-frequency normal modes to generate the anisotropic variance for each residue from a large number of modes, weighted by the calculated frequencies, and compare the result with the crystallographic variation. This method has been used in previous work applying normal modes to crystallographic refinement \cite{kidera1990-3718}, but is not in common use for validating normal modes with experimental displacements. Instead, the procedure is often to project low-frequency modes individually onto a conformational change, and to obtain a cumulative projection coefficient. This, however, is impossible to do without prior knowledge of the conformational change in the structure, and gives only an agreement between the subspace spanned by a several modes and the conformational change.  Our approach does not presume any knowledge beyond the initial structure, and measures agreement with the entire normal mode ensemble, rather than with individual modes.

This is also, as far as we know, the first large-scale comparative study of coarse-grained normal mode methods. Comparison of the modes from different potentials reveals a distinct split between the ENM methods and BNM, as seen in Figure \ref{fig:mode_comp}.  This suggests that the chemical information which is ignored by the ENM approach is observable in the BNM results, although there is significant similarity at low frequency modes due to the shape of the structure reflected in both potential types.  The observation opens up a possibility of separating the effect of gross protein structure from that of detailed residue chemistry as reflected by the CHARMM forcefield. A careful comparison of ENM predictions with those from normal modes with chemical forcefield could potentially be used to determine residues whose chemistry plays a key role in the dynamic coupling in the structure, and which would therefore be especially sensitive to mutation. The visual comparison of the correlation patterns from ANM and BNM demonstrates that the chemical effects are subtle in comparison to the topological features captured by both BNM and ANM, and all the other methods.

The choice of computational strategy to address a given problem involves balancing computational efficiency against model detail.  Fast calculations are meaningless if they give unreliable results, and extremely accurate calculations are of no use if they cannot be completed in a reasonable time frame. Normal mode analysis is based on a choice to limit the model to the neighborhood of the potential minimum. Further simplification of using an elastic network model potential instead of a physical, all-atom potential is another concession towards efficient calculation and away from physical reality. BNM prediction quality of the magnitude of flexibility is similar or worse than those from ENM models. This once again demonstrates the robustness of the elastic network models, and suggests that the main factor in determining macromolecular flexibility is the number of local contacts, determined by the shape of the molecule \cite{halle2002-1274}.  In prediction of directionality of motion, there is a clear difference between methods that are based only on C$\alpha$ coordinates, (HCA and ANM) and those that consider all atoms. CHARMM-based BNM has the best directional agreement as measured by ellipsoid correlation, while our new method, DNM, and ElN come close to matching this standard. This suggests that an all-atom ENM potential can give an accurate representation of the conformational ensemble of a protein near the native state, but the inclusion of chemical forces improves the model.

We must also consider the cost, both computational and human, required by the different methods.  One of the main differences between elastic network model techniques and BNM is that the latter requires an initial minimization step (see Methods).  If minimization is not complete, subsequent diagonalization will lead to spurious modes with large, negative frequencies; one must be careful to only pick productive modes when using results BNM, while elastic network models are at a local minimum by construction.  Further, the initial setup with an all-atom potential requires attention to the individual oddities of each structure: disulfide bonds, non-standard residues, bound ligands or cofactors.  Each of these issues must be dealt with individually, thus making automation of the calculations more difficult. Compared with ENM models, in which most of these details are ignored, CHARMM-based normal modes require a great deal of human effort.

The present results indicate that anisotropic temperature factors from high resolution crystal structures contain a measure of internal molecular flexibility, and can be used as a source of dynamic information and as a test for computational methodologies. Comparison of different methods indicates that elastic network models can describe the conformational ensemble of protein structures with accuracy approaching that of CHARMM, but that there is a substantial spread in prediction quality of different ENM potentials. Using an exclusively C$\alpha$-based potential results in a large sacrifice in prediction quality of directionality, but the lowest frequency modes are robust across the methods. The information may help those studying interactions within biological molecules choose the appropriate level of complexity for the system of interest and for the level of detail required of the prediction.

\section*{Computational procedures}
We use normal mode analysis \cite{go1983-3696,brooks1983-6571,levitt1985-423} to predict the positional ensemble of protein structures. The different models use distinct potentials, all of which require the knowledge of protein structure.  The Hessian matrix of the potential is diagonalized to find the normal modes, or eigenvectors $u_i$ and the corresponding frequencies $\omega_i$: $Hu_i = \omega_i^2 u_i$. The decomposition allows us to compute the covariance matrix, which is proportional to the pseudo-inverse of the Hessian.  Let $\delta_i$ be the deviation from the mean for component $i$, then the covariance between two deviations is:

\begin{equation}
<\delta_{i}\delta_{j}> = \frac{1}{2k_B T}\sum_k\frac{1}{\omega_k^2} u_{ik}u_{jk}
\label{eq:msf}
\end{equation}
where $u_{ik}$ is the $i$th component of the $k$th normal mode with frequency $\omega_k$. Note that the modes with the lowest frequencies make the greatest contribution to residue mobility, so a small fraction of all the modes is sufficient to obtain a good approximation of the sum. This allows us to compute anisotropic variances as 3x3 blocks around the diagonal of the covariance matrix \cite{kidera1990-3718}.

We may also compute the correlation coefficient  between the deviations of any two atoms, to generate the global correlation matrix:
\begin{equation}
R(\delta_{i}, \delta_{j}) = \frac{<\delta_{i}\delta_{j}>}{\sqrt{<\delta_{i}\delta_{i}><\delta_{j}\delta_{j}>}}
\end{equation}

\subsection*{Elastic Network Models} Anisotropic Network Model (ANM) \cite{atilgan2001-505} is a version of Elastic Network Model (ENM), based on connecting residues with C$\alpha$ atoms within a cutoff distance $R_c$ with spring-like interactions. The Hessian matrix is a 3Nx3N matrix, where N is the number of residues, consisting of 3x3 submatrices $H_{ij}$ which depend on the direction of the vector between C$\alpha$ atoms $i$ and $j$, and are 0 if the C$\alpha$s are more than $R_c$ apart.  The diagonal submatrices are defined as follows: $H_{ii} = - \sum_j H_{ij}$. This defines a coarse-grained Elastic Network Model of a protein structure with directional information.  We implemented this alrogithm using perl code to read PDB files and construct the Hessian, with MATLAB  (The Mathworks, Inc., Natick, MA) scripts used for diagonalization.

We introduce two modifications to ANM, analogous to those we had previously proposed for isotropic models \cite{kondrashov2006-2760}, and term the new model Distance Network Model (DNM).  First, the connectivity of the elastic potential is based on distances between nonhydrogen atoms of residue pairs, instead of only the C$\alpha$ atoms. The contacts from all atoms are combined for each residue to yield an interaction potential at the residue level. Second, we introduce different classes of residue interactions based on interatomic distances, with distinct Hookean spring constants. We use distance bins to define the interaction classes, specifically, covalent interactions are found by distance less than 2.3 \AA, the next shell is up to 3.3 \AA, followed by 5, 7, 9, and 11 \AA.  The Hessian matrix for each bin is defined exactly as for ANM above, with the difference that the equilibrium distance between two atoms has to be in the distance bin, while the coordinates $(x_i,y_i,z_i)$ for residue $i$ remain the C$\alpha$ coordinates. If $H_a$ is the contact matrix for class $a$, the total Hessian matrix for DNM is a linear combination of the matrices, with $k_a$ as the interaction constant for each class: 
\begin{equation}
H_{total} = \sum_a k_a H_a =  \sum_a \frac{H_{a}}{tr(H_{a})}
\label{eq:sum_hess}
\end{equation}
The constants $k_a$ define the strength of interactions, and we choose to use the total number of contacts in each class as a normalization constant, $k_a = 1/tr(H_a)$.  Thus, DNM adds several different interaction constants, but these are defined from the contact matrices, and thus are not free parameters to be optimized. The only free parameter, as in other ENM, is the cutoff distance for atomic contacts, which we vary from 5 to 11 \AA, as described in Results. The implementation again used a combination of perl and MATLAB scripts.

The details of the normal mode analysis implementation of Molecular Modeling ToolKit have been described elsewhere \cite{hinsen2000-79,hinsen1999-369}.  For this study, we used the harmonic C$\alpha$ forcefield (HCA) \cite{hinsen2000-25}, which defines different interaction constants for covalently bonded and non-covalently bonded C$\alpha$ atoms. The model uses the reciprocal of distance  to weight the harmonic interaction constants, and no parameters are varied from the default values. The MMTK calculations for all 83 structures in the dataset are carried out in 2 hours on a single 2 GHz AMD Athlon processor with 2 GB of RAM. 

ElNemo (ElN) \cite{suhre2004-W610}, is an all-atom ENM, which constructs a contact matrix for all atoms within a certain radius, and then treats blocks of one or more residues, as rigid bodies using the Rotation-Translation blocking algorithm \cite{tama2000-1}.  The two main programs that constitute ElNemo, pdbmat and diagrtb, were kindly provided by the authors and installed on the local cluster.  All blocking is done on a residue by residue basis and  the interaction cutoff distance is varied from 4 \AA\ to 11 \AA.  Running ElN on all 83 structures in the dataset using 8 different cutoff distances took roughly 1 day to complete on a 100 node cluster of 2.2 GHz Apple G5 processors with 4 GB of RAM.  

\subsection*{Block Normal Modes with CHARMM} Block normal-mode analysis (BNM), originally suggested by Tama, \emph{et. al.} \cite{tama2000-1} and subsequently improved by Li and Cui \cite{li2002-2457}, computes an all-atom Hessian which is then projected onto a blocked space spanned by the rotational and vibrational degrees of freedom of predefined blocks; in this work each residue is treated as a rotation-translation block, as in ElNemo method above. For this level of coarse-graining, the procedure reduces the Hessian storage space by approximately a factor of 25 and the diagonalization time by a factor of 125. The resulting blocked eigenvectors are then projected back to the all-atom space to give all-atom eigenvectors. This procedure perturbs the magnitudes of the eigenvalues, but in a linear fashion for the low-frequency modes \cite{tama2000-1,li2002-2457}. The appropriate scale factor of 1.7 has been used in this work. Local minimization is performed to ensure that the linear term in the Taylor expansion of the potential is zero \cite{hayward2001-1}.  This minimization is completed using cycles of the adapted-basis Newton-Raphson method with gradually decreasing harmonic constraints to remove local steric clashes without perturbing the structure significantly.  A final minimization with no harmonic constraints is performed until the RMS energy gradient reached 0.01 kcal/mol/\AA.  The average minimization time for this set is approximately eight minutes, but minimization times vary widely because of protein size: 54 seconds for the 52 residue 1RB9, and 73 minutes for the 325 residue 1O7J, all computed on 1.8 GHz Athlon single-processor station with 1 GB of memory, running Red Hat Linux 7.2. In some of the systems studied, this level of minimization resulted in modes with large negative frequencies in addition to the normal six rotational/translational modes.  In these cases, these modes are ignored for all subsequent calculations.  The average diagonalization time for BNM is approximately six minutes, varying widely again: 48 seconds for the 1RB9, and 51 minutes for 1O7J.   All calculations are completed using the CHARMM suite of programs \cite{brooks1983-187}.  The extended atom CHARMM19 force field \cite{neria1996-1902} modified for use with the EEF1 solvation model \cite{lazaridis1999-133} is used for both minimizations and the BNM.

\subsection*{Measures of agreement with crystallographic data} The data set is obtained by searching the Protein Data Bank \cite{berman2000-235} for protein structures determined by X-ray crystallography to at least 1.0 \AA\ resolution, containing at least 50 residues in a single chain.  Structures with more than 50 \% identity are discarded, leaving 98 non-redundant proteins, of which 87 contained ANISOU cards (anisotropic temperature factors); 4 more structures are discarded because they contained modified protein residues for which CHARMM libraries are not available.  The resultant set is structurally diverse, with all major SCOP superfamilies \cite{murzin1995-536} represented, as shown in Table 1 in Supplemental Materials.  All protein chains in the PDB files are kept in the model in order to best represent the crystal environment.  Copies of the protein molecule surrounding the structure in the crystal are generated using the symexp command in PyMOL \cite{pymol-0}, and residues with at least one atom less than 4 \AA\ from an atom in a crystal copy are considered to be involved in crystal contacts.  These residues, along with those with C$\alpha$ atoms with occupancy less than 1, are not used in ADP comparisons. 

The anisotropic parameters are 3x3 matrices that define the variance of a 3-dimensional Gaussian probability distribution for position of each atom: 
$$ \rho(\vec R) = \left( \frac{\det U^{-1}}{3 \pi^3}\right)  exp\left(-\frac{1}{2}\vec R^{T}  \left(\begin{array}{ccc}U_{xx} & U_{xy} & U_{xz} \\U_{xy} & U_{yy} & U_{yz} \\U_{xz} & U_{yz} & U_{zz}\end{array}\right)^{-1} \vec R \right) $$ 

The six components of ADPs, $U_{xx}$, etc., are reported in PDB files in ANISOU cards \cite{berman2000-235}. We compare the computationally predicted anisotropic parameters $V$ with those from the crystal structures $U$.  Quality prediction of magnitude of motion is measured by linear correlation of the traces of the matrices $U$ and $V$ over the whole structure, which we call isotropic correlation (IC). To compare directions of ellipsoids, we first divide all the matrices by their trace, to set all magnitudes to 1.  Ellipsoids are described by their principal axes (eigenvectors) and the associated lengths (inverse eigenvalues); the ratio of the smallest to the largest eigenvalue is called its anisotropy \cite{trueblood1996-770}.  We restrict directionality comparison to ellipsoids with anisotropy of less than 0.5, since directional comparison of near-spherical ellipsoids is meaningless. The simplest comparison of directionality is the absolute value of the dot product between the major directions. It is a rough estimate of agreement for two ellipsoids whose major axes are dominant, but has the virtue of simplicity.  A more systematic measure of ellipsoid similarity was proposed by Merritt \cite{merritt1999-1997}, based on computation of the overlap integral between two probability densities. This measure, known to crystallographers as real-space correlation coefficient, is defined for two three-dimensional Gaussian distributions with covariance matrices $U$ and $V$ as follows:
\begin{equation}
cc(U,V) = \frac{(\det (U^{-1})\det (V^{-1}))^{1/4}}{\left[1/8 \det(U^{-1} + V^{-1})\right]^{1/2}}
\end{equation}

We also compare modes produced by the different normal mode potentials. To compare mode $i$ (as ordered by frequency) from two methods, we take the average between the best agreement for mode $i$ from method $a$ with modes from method $b$, and the best agreement for mode $i$ from method $b$ with the modes from method $a$.  We compare the modes similar in frequency ordering, specifically, only the modes no more than 3 indices higher or lower. The formula for overlap for mode $i$ between method a and b is:
\begin{equation}
O_{a,b}(i) = \frac{1}{2}\max_{i-3\le j \le i+3} \vec u^a_i \cdot \vec u^b_j + \frac{1}{2}\max_{i-3\le j \le i+3}  \vec u^b_i \cdot \vec u^a_j 
\end{equation}
If the best agreement is between modes of the same index, then the two maxima are the same. Figures \ref{fig:mode_comp} and \ref{fig:corr_mat} were prepared using MATLAB (The Mathworks, Inc., Natick, MA) and figure \ref{fig:pdz_anis} with rastep and raster3d \cite{merritt1997-505}.

\section*{Acknowledgments}
We thank Karsten Suhre and Yves-Henri Sanejouand for providing the source code for ElNemo software. D.A.K. and R.M.B. were supported through an National Library of Medicine training grant to the Computation and Informatics in Biology and Medicine program at UW-Madison (NLM 5T15LM007359), with R.M.B. also receiving support from a training grant from the Department of Energy, Genomes to Life project (DE-FG2-04ER25627). A.W.VW. was supported by an NSF pre-doctoral fellowship. Q.C. is an Alfred P. Sloan Research Fellow.

\bibliography{bibkon}

\newpage
\section*{Tables}
\begin{table}[htbp]
\begin{centering}
   \caption{Prediction quality using different NMA potentials} 
   \begin{tabular}{@{\vrule height 10.5pt depth4pt  width0pt}lccc} 
  			& dot$^*$	& real space$^*$ & isotropic corr$^\dagger$ \\ 
        \hline
        random	& 0.5			&	0.3			&  0	\\
        HCA   	& 0.599 (0.199)	& 0.520 (0.167)	& 0.617 (0.131) \\
        ANM  	& 0.556 (0.190) 	& 0.525 (0.172)	& 0.676  (0.111)\\ 
        DNM		& 0.650 (0.209)	& 0.575 (0.181)	& 0.655 (0.136)  \\    
        	ElN		& 0.641 (0.208)	& 0.583 (0.184)	& 0.680 (0.128) \\
	BNM		& 0.658 (0.211)	& 0.608 (0.188)	& 0.658 (0.128) \\
   \end{tabular} \\
 $^*$ mean and standard deviation over all individual ADPs;  $^\dagger$ mean and standard deviation over 83 structures.
   \label{tab:results_summary}
   \end{centering}
\end{table}

\newpage
\section*{Figures}

\end{document}